\documentclass[11pt,twoside]{article}

\usepackage{asp2006}
\usepackage{epsf}
\usepackage{psfig}
\usepackage{lscape}

\begin{document}
\title{Phase-referencing on BL~Lac}   
\author{A.~Sokolov and T.~Cawthorne}   
\affil{University of Central Lancashire}    

\begin{abstract} 
We report the results of a phase-referencing study aimed at uncovering precession of the VLBI jet of BL Lac. The
observations were conducted at 8, 15, 22, and 43 GHz and consist of seven epochs spanning about two years. We
investigated the change in the absolute position of BL Lac's radio core by means of phase-referencing with two nearby
sources, 2151+431 and 2207+374. The shift in the position of the core perpendicular to the jet is a signature of precession.
However, the periodic variations with an amplitude of $\sim$0.15 mas and a period of 1 year can be attributed to seasonal weather
variations. We also detect a trend in position of the core on the scale of $\sim$0.1 mas over two years.
\end{abstract}


\begin{figure}[!ht]
\plotone{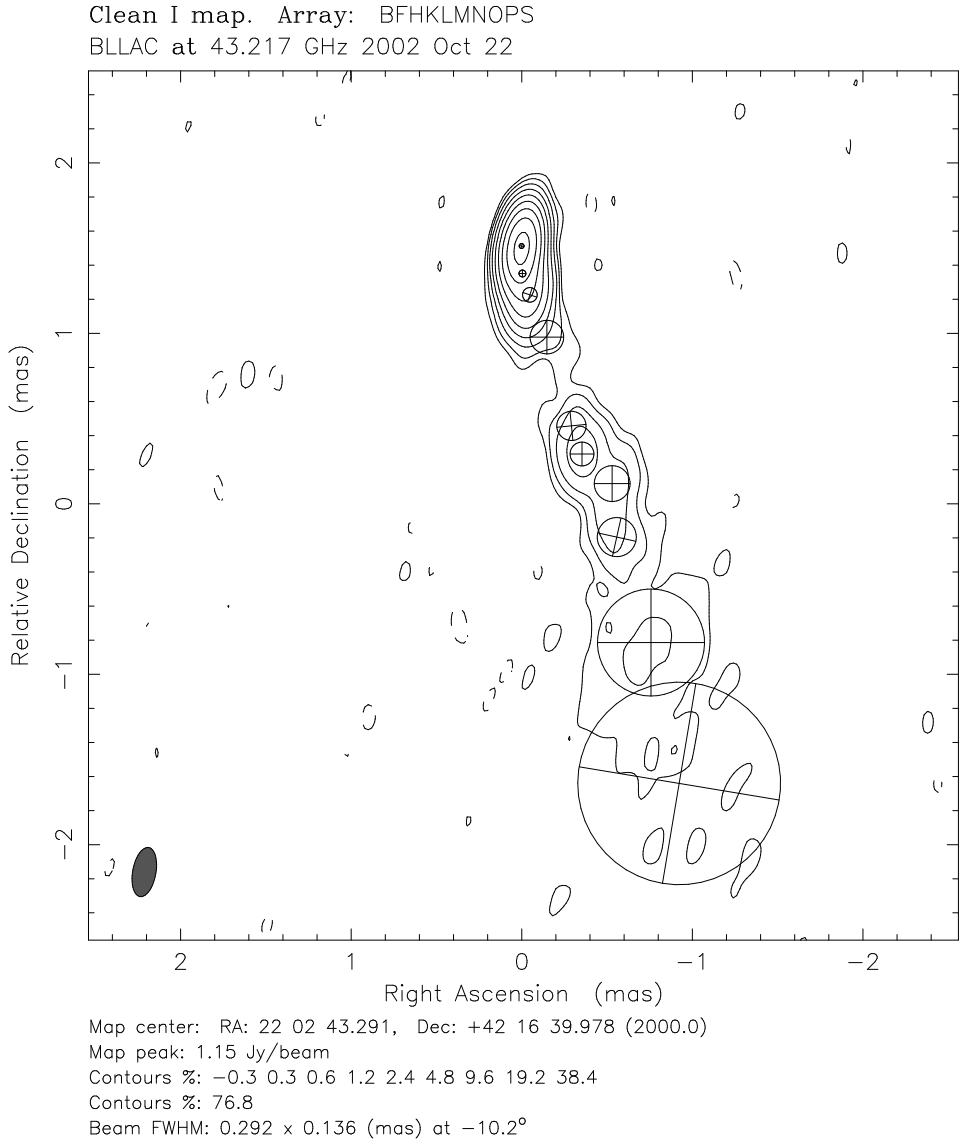}
\label{bllac}
\caption{An image of BL~Lac at 43 GHz for the first epoch in our observing program.
The phase-reference centre is shifted along RA by 1.5~mas for convenience; uniform weighting is used. 
The emission is fitted with ten circular components of various brightness and size.
The upmost model component in the image was used as a reference point for tracking variations in
position of the phase-reference sources.}
\end{figure}

\begin{figure}[!ht]
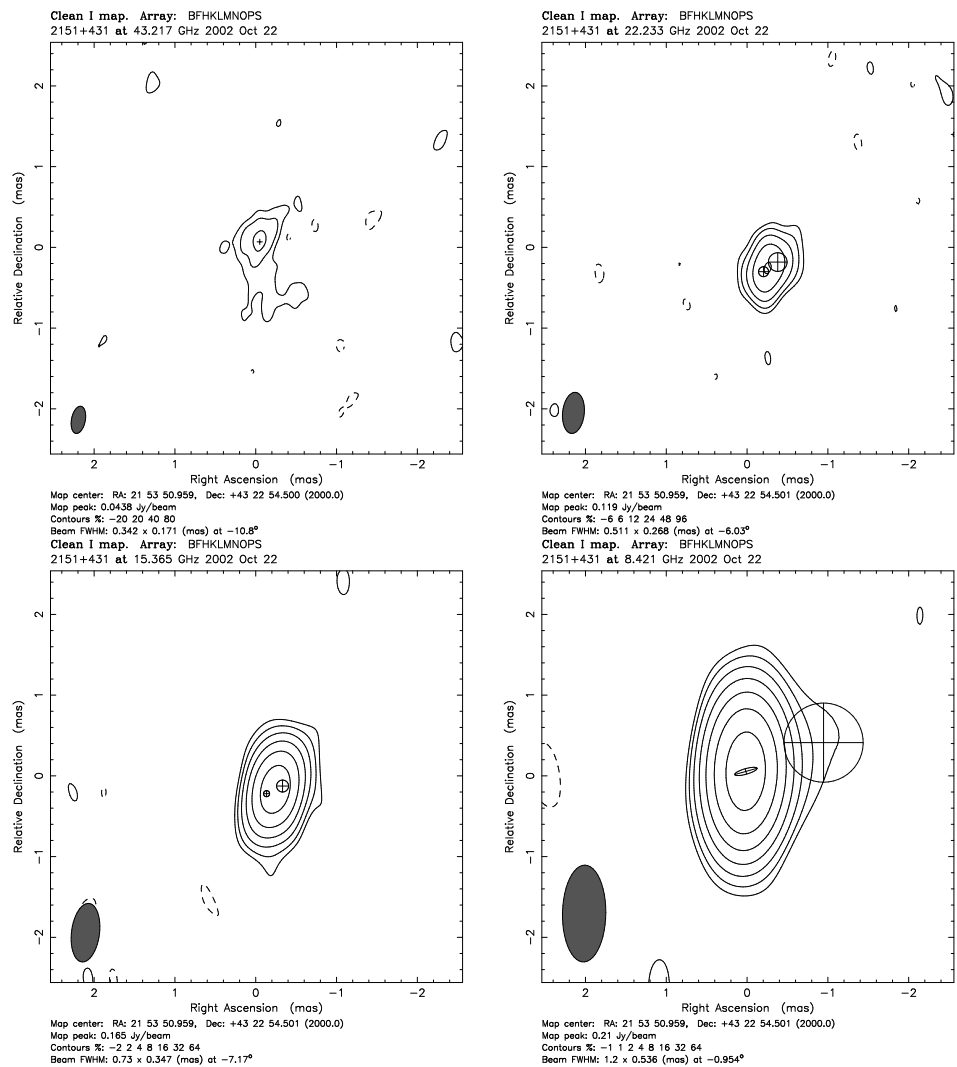

\label{2151+431}
\plottwo{2151+431.43u.eps}{2151+431.22.eps}
\plottwo{2151+431.15.eps}{2151+431.08.eps}
\caption{Four images of 2151+431 at 43, 22, 15, and 8~GHz for the same epoch as in Figure~1.
Natural weighting is used for the 43~GHz image, uniform weighting is used for the other frequencies.
Circular components represent the model fit at 22 and 15~GHz. The 43~GHz image was fitted by a point-like component instead of a usual gaussian.
A single elliptical component was used to fit the peak emission at 8~GHz. 
An additional circular component at 8~GHz is weak and does not affect the position of the main elliptical component.
}
\end{figure}

The appearance of the superluminal components in BL Lac suggests that the direction of the jet 
changes periodically within about 30 degrees in projection. Recent studies by \citet{stirling}
confirm these findings with several independent indicators. These include the changes in the structural position angle of the jet
at the radio core as well as the changes in the orientation of the polarised component of emission. The data covering four years
were analysed and revealed periodic changes in these parameters with the period of 2.3~years. 
However, it was not clear if this pattern would repeat in the future (Mutel \& Denn, 2005). 
It is possible that these changes were due to an instability in the jet rather than precession. 
Further studies were required to establish if it is indeed the precession that was observed.

To advance this work further and to establish directly if the changes in the jet direction of BL~Lac are indeed caused by precession
we have carried out an additional observational campaign of BL~Lac. 
In this campaign we utilized the technique of phase-referencing in order to observe directly the changes in the
absolute position of BL~Lac's core as a function of time. 
Such changes could provide direct evidence for precession of the parsec-scale jet in BL~Lac. 
Indeed, since the radio core is located at some distance away from where the jet is produced,
precession of the jet should cause the absolute position of the core to wobble with time
in the direction normal to the jet. Moreover, these changes should be regular and periodic.

During the years 2002--2004 we obtained seven observations of BL Lac at 8, 15, 22, and 43~GHz.
Two nearby sources, 2151+431 and 2207+374, were observed concurrently for the purpose of phase-referencing.
These are separated from BL Lac by 2.5$^\circ$ and 5$^\circ$ respectively.
The cycle time for phase-referencing was $\sim90$ seconds at 43 GHz, consisting of $50$ seconds on BL Lac and $8$ seconds on 2151+431
 (2207+374 was not observed at 43 GHz). Somewhat higher cycle times were used at lower frequencies. 
Initially the project aimed at observing BL Lac during a single year, once every two months.
However, in the end seven observations were carried out, spanning almost two years, with epoch 2 and epoch 3 separated by
over 7~months.

We reduced the data in AIPS following standard recipes for the calibration of the VLBA data.  
The EOP corrections have been applied to all epochs at the first step of the calibration (VLBA test memo 69).
The signal-to-noise ratio for the phase-reference sources, 2151+431 and 2207+374, was too poor to allow a stable fringe solution.
Therefore, we used BL~Lac as our calibrator for phase-referencing, while the phase-reference sources played the role of the targets.
For the phase-referencing experiments it is usual to obtain a fringe solution for the calibrator with the AIPS command FRING.
Then the solution can be applied to the phase-reference sources,
thus preserving the information about their spatial separation from the calibrator (see AIPS Cookbook).
However, we avoided this route and instead adopted the following procedure (S.~Jorstad, private communication).
A fringe solution for the calibrator, i.e.~BL~Lac, was derived and applied to the calibrator only.
After this the calibrator was exported from AIPS and imaged in Difmap. The fringe solution obtained in the previous step was discarded.
The clean model derived in Difmap was imported back into AIPS.
This clean model was then used in the AIPS command CALIB to find a phase solution for the calibrator.
This solution was then applied to both the calibrator and the phase-reference sources.
Finally, all sources were imaged in Difmap and the model fits using circular components were determined.
Some of the representative images are shown in Figures~1 and~2.

The image of BL~Lac at 43~GHz for the first epoch is shown in Figure~1. 
Several model components correspond to the radio core. 
Unfortunately, due to insufficient north-south uv coverage (observation was awarded only 6~hours per epoch)
we failed to resolve the fine structure of BL~Lac's radio core, which was previously found to comprise two components (C1 and C2 in \citet{stirling}).
Despite this caveat we selected the most upstream component as a reference point.
The images of 2151+431 at all four observed frequencies for the same epoch are shown in Figure 2.
The position of the source in the original images is displaced from the centre of the map by $\sim$8~mas
due to phase-referencing.
This shift represents the discrepancy in the assumed and observed position of 2151+431.
However, for convenience we introduced a constant shift of position in the displayed images
since we are only interested in the change of the position with time rather than the absolute value of the position at each epoch.
The source 2207+374 was found to have a rather complex extended structure consisting of several components,
separated by as much as 5~mas. A reliable model of the emission could only be established at 8~GHz.
For these and other reasons this source produced inconsistent results and it will not be discussed in the present paper.

We found two separate components in 2151+431 at both 15 and 22 GHz and an indication of the third faint component at 8~GHz for some epochs.
It appears that there is a jet in 2151+431 oriented roughly westward.
Consistent with the jet scenario we found that the western component exhibits a rather high spectral index 
indicative of a shock wave moving downstream. The other component exhibits a harder spectrum.
Further analysis suggested that the separation between these two components is increasing with time.
We decided to use the component with the hard spectrum as a reference point for our studies.
The S/N ratio at 43~GHz was too poor to detect the two-component structure revealed at lower frequencies.
The emission was fitted with a single point-like component whose position was used as a reference point.
The resolution at 8~GHz was insufficient for detecting this structure and a single elliptical component was used
to fit most of the emission. The centre of the elliptical component was used as a reference.

\begin{figure}[!ht]
\plottwo{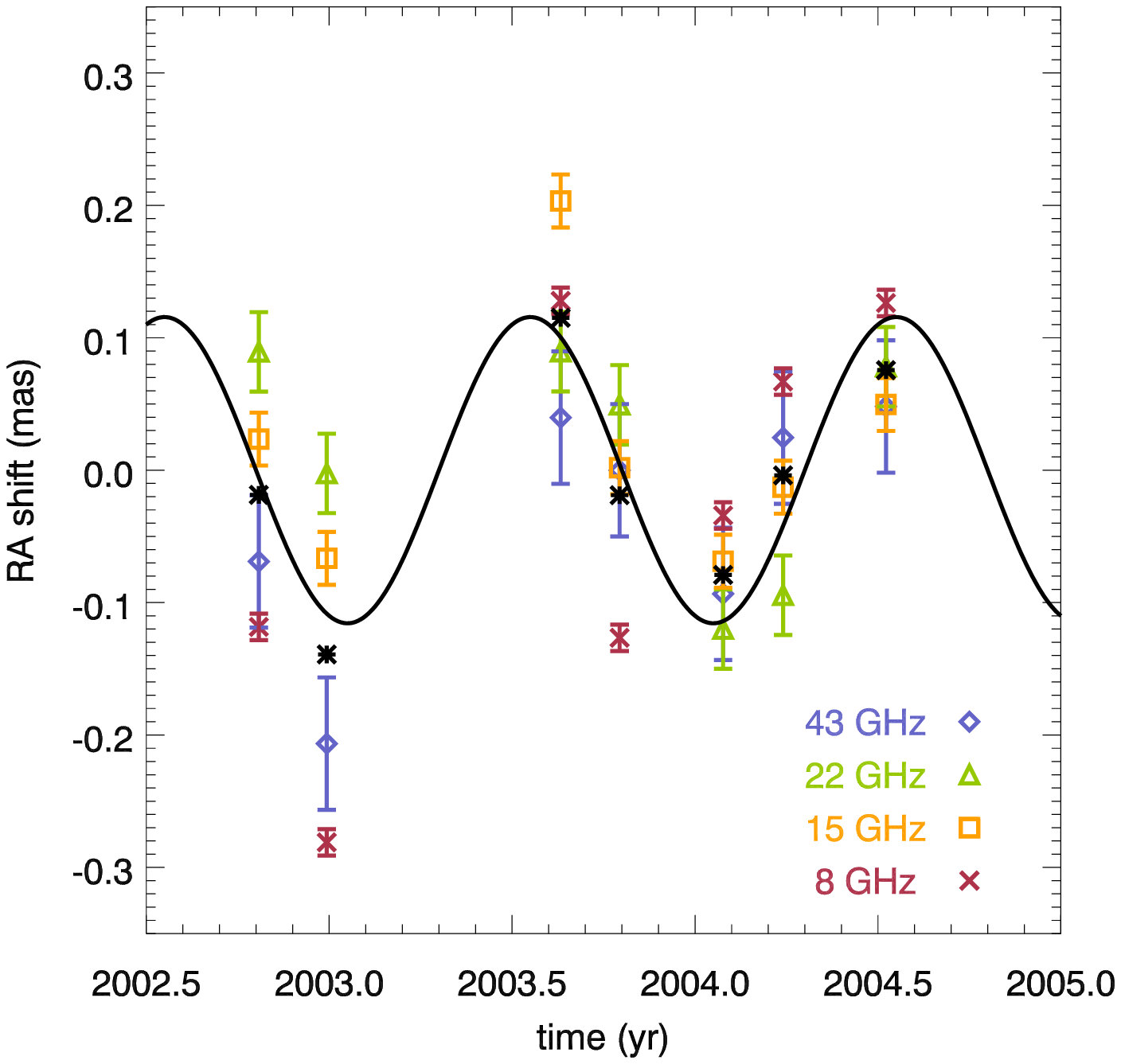}{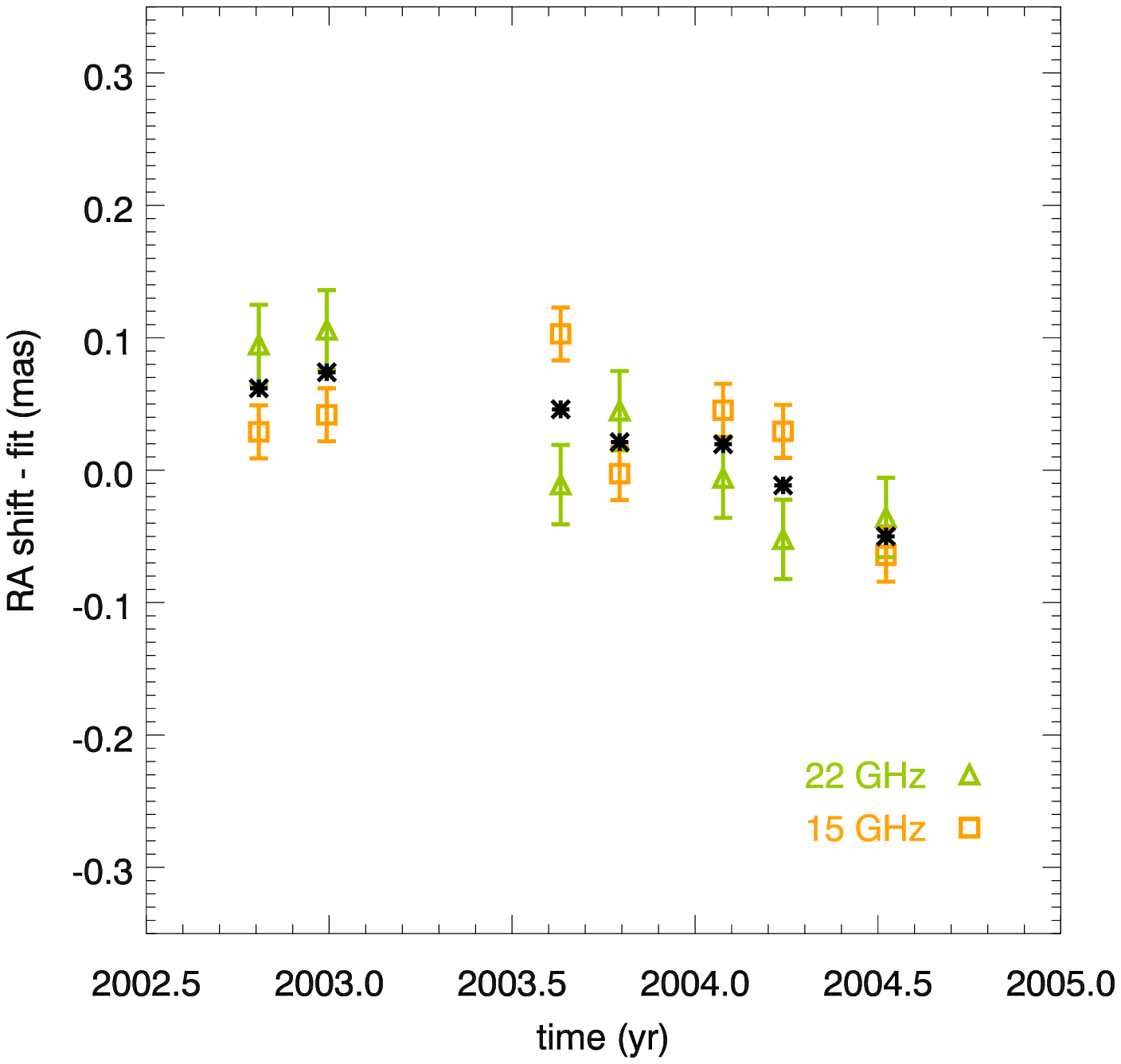}
\label{shift}
\caption{
Left: the shift in the position of BL Lac's core, i.e.~reference model component, along right ascension 
inferred from the measurements of the position of 2151+431 for all seven epochs in the observing program. 
The error-bars are derived from the beam width at the corresponding frequency divided by the S/N ratio of the 
reference model component in 2151+431.
Note that the error-bars for the 8~GHz data are likely to be much larger than the ones shown 
since the position of a single elliptical component is likely to be affected by both features observed at higher frequencies.
The sinusoidal fit is shown in black; asterisks represent the averages over frequencies for each epoch.
Right: the same data after subtracting the sinusoidal fit, only 22 and 15~GHz data are shown.
}
\end{figure}

The results of our analysis are shown in Figure~3. 
The reference component in 2151+431 was assumed to be fixed in position on the sky.
The shift refers to the inferred displacement in the absolute position of the reference component of BL Lac.
We found a variation in the position of BL Lac's core with the period of 1~year and an amplitude of $\sim$0.15~mas,
in the direction normal to the jet.
It is likely that the observed variation is due to the errors in the clock and/or atmospheric component of the correlator model.
It was not possible to remove these errors with DELZN (Mioduszewski, AIPS memo 110) since the calibrators necessary
for this procedure were not observed. 
Another recent AIPS procedure ATMCA (Fomalont \& Kogan, AIPS memo 111) was employed but it failed to provide a reliable 
solution due to low S/N of the phase-reference sources.

Analysis of the residuals after the sinusoidal fit had been removed revealed a slow trend at 15 and 22 GHz,
with the position of the core changing by $\sim$0.1~mas  over a period of two years. 
The 8 and 43 GHz data needed to be rejected for this trend to become visible since the the S/N is too poor at 43 GHz
while the beam size is too great at 8 GHz. It is not clear whether this trend is due to processes in BL Lac itself.
Since it was not possible to remove the effect of the atmosphere properly, this trend could be due to some long-term weather variations.
It could also be due to some changes in the phase-reference source, 2151+431, whose jet appears to be oriented in the direction normal to the jet in BL Lac.
To address these concerns further phase-reference observations of BL Lac are necessary.
These observations must take into account the atmospheric and clock errors in the correlator model.



\end{document}